# Fluctuation-electromagnetic interaction of a small particle with evanescent modes of the surface: impact of translational motion and rotation

A.A. Kyasov and G.V. Dedkov

Nanoscale Physics Group, Kabardino-Balkarian State University, Nalchik, 360004, Russia

Based on the fluctuation-electromagnetic theory, we have calculated tangential (frictional) and normal forces, radiation heat flux and frictional torque on a small rotating particle moving at a nonrelativistic velocity close to a smooth featureless surface. The particle and surface are characterized by different temperatures corresponding to the local thermodynamic equilibrium in their own reference frames.

.

## 1. Introduction

Fluctuation electromagnetic interaction between a small particle in uniform translational motion and a polarizable surface is the subject of continuing interest [1--3]. Very often, the corresponding problems have been addressed by several authors using different approaches, that has led to many contradictions. For a detailed discussion of these issues see [4--6]. Quite recently, using a fully covariant framework [7], Pieplow and Henkel have recovered our basic relativistic results [4,8] for the radiative friction forces in the case of a particle motion through a homogeneous radiation background and above a planar interface. These results also agree with [9] and other well-established nonrelativistic results. Thus, one can assert that the theory [4--9] is based on a unified basis and thus leads to mutually concordant results.

A new direction of studies has been triggered by Manjavacas and Garcia de Abajo [10], who calculated frictional torque and thermal radiation intensity on a particle rotating in vacuum. Further on, we and other authors have calculated frictional torque, thermal radiation intensity and the force of attraction on a particle rotating near the surface [11,12]. For a compact review of these and some other recent results of reference character see in [13].

The aim of this paper is to consider a combined action of translational motion and particle rotation near a polarizable surface within the nonrelativistic approximation. We consider several directions of the rotation axis with respect to the velocity and obtain closed analytical expressions for the tangential/normal forces, intensity of thermal radiation and frictional torque acting on the particle. All previous results in the case of translational motion without rotation and rotation without translational motion follow from the obtained formulas.

## 2. Theory

Following our general method [4,8], consider a small spherical particle of radius $R$ with electric polarizability $\alpha(\omega)$ and temperature $T_1$, moving with nonrelativistic velocity $V \ll c$ ($\mathbf{V}=(V,0,0)$) parallel to a smooth featureless surface with dielectric permittivity $\varepsilon(\omega)$ and temperature $T_2$. The boundary $z=0$ separates the upper vacuum space from the dielectric medium. Simultaneously, the particle rotates with the angular velocity $\mathbf{\Omega}$, the direction of which is assumed to be $\mathbf{\Omega}=(\Omega,0,0)$ (Fig.1a), $\mathbf{\Omega}=(0,\Omega,0)$ (Fig.1b), and $\mathbf{\Omega}=(0,0,\Omega)$ (Fig.1c). So, the system is assumed to be stationary, but out of thermal and dynamical equilibrium. Assuming the conditions $R \ll z_0, R \ll \min\{2\pi c/\omega_0, 2\pi c/\Omega, 2\pi\hbar c/k_B T_1, 2\pi\hbar c/k_B T_2\}$ to be fulfilled, one can consider the particle as a point-like fluctuating dipole with neglect of retardation.

Let us consider, for definiteness, a configuration shown in Fig. 1a. In this case, we have three coordinate systems, namely $\Sigma, \Sigma', \Sigma''$. System $\Sigma'$ moves with velocity $\mathbf{V}$ relative to the frame $\Sigma$ that is rigidly connected to the surface, while the velocity $\mathbf{\Omega}$ is given in $\Sigma'$. From the very beginning, the values of $\alpha(\omega)$ and $T_1$ are defined in the rest frame $\Sigma''$ of the particle rotating with velocity $\mathbf{\Omega}$ relative to $\Sigma'$.

Within the nonrelativistic approximation used in this work, several physical values prove to be the same in all reference systems $\Sigma, \Sigma', \Sigma''$. However, bearing in mind future generalization of these results in relativistic case, designating initial reference frame for each physical value is expedient from the very start.

The starting equations for the tangential/normal forces $F_{x,z}$, the frictional torque $M_{x,y,z}$ and the particle heating rate $\dot{Q}$ are given by

$$F_x = \left\langle \nabla_x \left( \mathbf{d}^{sp} \cdot \mathbf{E}^{ind} + \mathbf{d}^{ind} \cdot \mathbf{E}^{sp} \right) \right\rangle \equiv F_x^{(1)} + F_x^{(2)} \tag{1}$$

$$F_z = \left\langle \nabla_z \left( \mathbf{d}^{sp} \cdot \mathbf{E}^{ind} + \mathbf{d}^{ind} \cdot \mathbf{E}^{sp} \right) \right\rangle \equiv F_z^{(1)} + F_z^{(2)} \tag{2}$$

$$\dot{Q} = \left\langle \left( \dot{\mathbf{d}}^{sp} \cdot \mathbf{E}^{ind} + \dot{\mathbf{d}}^{ind} \cdot \mathbf{E}^{sp} \right) \right\rangle \equiv \dot{Q}^{(1)} + \dot{Q}^{(2)} \tag{3}$$

$$M_i = \left\langle \mathbf{d}^{sp} \times \mathbf{E}^{ind} + \mathbf{d}^{ind} \times \mathbf{E}^{sp} \right\rangle_i \equiv M_i^{(1)} + M_i^{(2)}, i = x, y, z \tag{4}$$

As usually, the brackets $\langle ... \rangle$ denote total quantum-statistical averaging, indexes "sp" and "ind" denote spontaneous and induced components of the field and dipole moment of the particle, and the points above the values---derivatives with respect to time.

To calculate the contribution of $\mathbf{d}^{sp}$ (the first terms in (1)—(4)), we should take into account that the fluctuation-dissipation relationships for the spontaneous dipole moment of the particle in its rest frame $\Sigma'$ take the form

$$\left\langle d^{sp}{}'_x(\omega')d^{sp}{}'_x(\omega)\right\rangle = 2\pi\hbar\delta(\omega+\omega')\alpha''(\omega)\coth(\hbar\omega/2k_BT_1) \tag{5}$$

$$\left\langle d^{sp}{}'_y(\omega')d^{sp}{}'_y(\omega)\right\rangle = \left\langle d^{sp}{}'_z(\omega')d^{sp}{}'_z(\omega)\right\rangle = \frac{1}{2}2\pi\hbar\delta(\omega+\omega')\cdot$$
$$\cdot\left[\alpha''(\omega_+)\coth(\hbar\omega_+/2k_BT_1)+\alpha''(\omega_-)\coth(\hbar\omega_-/2k_BT_1)\right] \tag{6}$$

$$\left\langle d^{sp}{}'_y(\omega')d^{sp}{}'_z(\omega)\right\rangle = -\left\langle d^{sp}{}'_z(\omega')d^{sp}{}'_y(\omega)\right\rangle = -\frac{i}{2}2\pi\hbar\delta(\omega+\omega')\cdot$$
$$\cdot\left[\alpha''(\omega_+)\coth(\hbar\omega_+/2k_BT_1)-\alpha''(\omega_-)\coth(\hbar\omega_-/2k_BT_1)\right] \tag{7}$$

where $\omega_\pm = \omega\pm\Omega$.

To calculate the contribution from $\mathbf{E}^{sp}$ corresponding to the particle rotation relative to the x-axis (the second terms in (1)—(4)), we should take into account relationships between the induced dipole moment of the particle and spontaneous field of the surface given in system $\Sigma$:

$$d^{ind}{}_x(t) = \int\frac{d\omega d^2k}{(2\pi)^3}\alpha(\omega-k_xV)E^{sp}{}_x(\omega,\mathbf{k})\exp(-i(\omega-k_xV)t) \tag{8}$$

$$d^{ind}{}_y(t) = \int\frac{d\omega d^2k}{(2\pi)^3}\frac{1}{2}\exp(-i(\omega-k_xV)t)\cdot$$
$$\cdot\left[\alpha(\omega-k_xV+\Omega)\left(E^{sp}{}_y(\omega,\mathbf{k})+iE^{sp}{}_z(\omega,\mathbf{k})\right)+\alpha(\omega-k_xV-\Omega)\left(E^{sp}{}_y(\omega,\mathbf{k})-iE^{sp}{}_z(\omega,\mathbf{k})\right)\right] \tag{9}$$

$$d^{ind}{}_z(t) = \int\frac{d\omega d^2k}{(2\pi)^3}\frac{1}{2}\exp(-i(\omega-k_xV)t)\cdot$$
$$\cdot\left[\alpha(\omega-k_xV+\Omega)\left(E^{sp}{}_z(\omega,\mathbf{k})-iE^{sp}{}_y(\omega,\mathbf{k})\right)+\alpha(\omega-k_xV-\Omega)\left(E^{sp}{}_z(\omega,\mathbf{k})+iE^{sp}{}_z(\omega,\mathbf{k})\right)\right] \tag{10}$$

In (8)—(10), the Fourier components of the fluctuating electric field of the surface over frequency $\omega$ and wave vector $\mathbf{k} = (k_x, k_y)$ are taken at the point $z = z_0$.

In configurations shown in Fig1b,c (rotation around the $y-$ and $z-$ axis) the corresponding fluctuation-dissipation relations and expressions for the dipole moments are obtained from (5)—(10) by a cyclic permutation $z \to x, x \to y, y \to z$. Using Eqs. (1)—(4) and Eqs. (5)—(10), the subsequent calculations are straightforward.

## 3. Results

### a) configuration $\Omega \parallel x$

$$F_x = -\frac{\hbar}{4\pi^2} \int_{-\infty}^{+\infty} d\omega \int d^2k \frac{k_x}{k} \exp(-2k z_0) \cdot$$
$$\cdot \{k_x^2 \Delta''(\omega)\alpha''(\omega+k_x V)[\coth(\hbar\omega/2k_B T_2) - \coth(\hbar(\omega+k_x V)/2k_B T_1)] +$$
$$(k_y^2 + k^2)\Delta''(\omega)\alpha''(\omega+k_x V+\Omega)[\coth(\hbar\omega/2k_B T_2) - \coth(\hbar(\omega+k_x V+\Omega)/2k_B T_1)]\}$$
(11)

$$F_z = -\frac{\hbar}{4\pi^2} \int_{-\infty}^{+\infty} d\omega \int d^2k \exp(-2k z_0) \cdot$$
$$\cdot \{k_x^2 \Delta'(\omega)\alpha''(\omega+k_x V)\coth(\hbar(\omega+k_x V)/2k_B T_1) + k_x^2 \Delta''(\omega)\alpha'(\omega+k_x V)\coth(\hbar\omega/2k_B T_2) +$$
$$+ (k_y^2 + k^2)\Delta'(\omega)\alpha''(\omega+k_x V+\Omega)\coth(\hbar(\omega+k_x V+\Omega)/2k_B T_1) +$$
$$+ (k_y^2 + k^2)\Delta''(\omega)\alpha'(\omega+k_x V+\Omega)\coth(\hbar\omega/2k_B T_2)\}$$
(12)

$$\dot{Q} = \frac{\hbar}{4\pi^2} \int_{-\infty}^{+\infty} d\omega \int d^2k \frac{\omega}{k} \exp(-2k z_0) \cdot$$
$$\cdot \{k_x^2 \Delta''(\omega)\alpha''(\omega+k_x V)[\coth(\hbar\omega/2k_B T_2) - \coth(\hbar(\omega+k_x V)/2k_B T_1)] +$$
$$(k_y^2 + k^2)\Delta''(\omega)\alpha''(\omega+k_x V+\Omega)[\coth(\hbar\omega/2k_B T_2) - \coth(\hbar(\omega+k_x V+\Omega)/2k_B T_1)]\}$$
(13)

$$M_x = -\frac{\hbar}{4\pi^2} \int_{-\infty}^{+\infty} d\omega \int d^2k \frac{(k_y^2 + k^2)}{k} \exp(-2k z_0) \cdot \Delta''(\omega)\alpha''(\omega+k_x V+\Omega) \cdot$$
$$\cdot [\coth(\hbar\omega/2k_B T_2) - \coth(\hbar(\omega+k_x V+\Omega)/2k_B T_1)]$$
(14)

### a) configuration $\Omega \parallel y$

$$F_x = -\frac{\hbar}{4\pi^2} \int_{-\infty}^{+\infty} d\omega \int d^2k \frac{k_x}{k} \exp(-2k z_0) \cdot$$
$$\cdot \{k_y^2 \Delta''(\omega)\alpha''(\omega+k_x V)[\coth(\hbar\omega/2k_B T_2) - \coth(\hbar(\omega+k_x V)/2k_B T_1)] +$$
$$(k_x^2 + k^2)\Delta''(\omega)\alpha''(\omega+k_x V+\Omega)[\coth(\hbar\omega/2k_B T_2) - \coth(\hbar(\omega+k_x V+\Omega)/2k_B T_1)]\}$$
(15)

$$F_z = -\frac{\hbar}{4\pi^2} \int_{-\infty}^{+\infty} d\omega \int d^2k \exp(-2k z_0) \cdot$$
$$\cdot \{k_y^2 \Delta'(\omega)\alpha''(\omega + k_x V) \coth(\hbar(\omega + k_x V)/2k_B T_1) + k_y^2 \Delta''(\omega)\alpha'(\omega + k_x V) \coth(\hbar\omega/2k_B T_2) + \quad (16)$$
$$+ (k_x^2 + k^2)\Delta'(\omega)\alpha''(\omega + k_x V + \Omega) \coth(\hbar(\omega + k_x V + \Omega)/2k_B T_1) +$$
$$+ (k_x^2 + k^2)\Delta''(\omega)\alpha'(\omega + k_x V + \Omega) \coth(\hbar\omega/2k_B T_2)\}$$

$$\dot{Q} = \frac{\hbar}{4\pi^2} \int_{-\infty}^{+\infty} d\omega \int d^2k \frac{\omega}{k} \exp(-2k z_0) \cdot$$
$$\cdot \{k_y^2 \Delta''(\omega)\alpha''(\omega + k_x V)[\coth(\hbar\omega/2k_B T_2) - \coth(\hbar(\omega + k_x V)/2k_B T_1)] + \quad (17)$$
$$(k_x^2 + k^2)\Delta''(\omega)\alpha''(\omega + k_x V + \Omega)[\coth(\hbar\omega/2k_B T_2) - \coth(\hbar(\omega + k_x V + \Omega)/2k_B T_1)]\}$$

$$M_y = -\frac{\hbar}{4\pi^2} \int_{-\infty}^{+\infty} d\omega \int d^2k \frac{(k_x^2 + k^2)}{k} \exp(-2k z_0) \cdot \Delta''(\omega)\alpha''(\omega + k_x V + \Omega) \cdot$$
$$\cdot [\coth(\hbar\omega/2k_B T_2) - \coth(\hbar(\omega + k_x V + \Omega)/2k_B T_1)] \quad (18)$$

**a) configuration $\Omega \parallel z$**

$$F_x = -\frac{\hbar}{4\pi^2} \int_{-\infty}^{+\infty} d\omega \int d^2k \frac{k_x}{k} \exp(-2k z_0) \cdot$$
$$\cdot \{k^2 \Delta''(\omega)\alpha''(\omega + k_x V)[\coth(\hbar\omega/2k_B T_2) - \coth(\hbar(\omega + k_x V)/2k_B T_1)] + \quad (19)$$
$$(k_x^2 + k_y^2)\Delta''(\omega)\alpha''(\omega + k_x V + \Omega)[\coth(\hbar\omega/2k_B T_2) - \coth(\hbar(\omega + k_x V + \Omega)/2k_B T_1)]\}$$

$$F_z = -\frac{\hbar}{4\pi^2} \int_{-\infty}^{+\infty} d\omega \int d^2k \exp(-2k z_0) \cdot$$
$$\cdot \{k^2 \Delta'(\omega)\alpha''(\omega + k_x V) \coth(\hbar(\omega + k_x V)/2k_B T_1) + k^2 \Delta''(\omega)\alpha'(\omega + k_x V) \coth(\hbar\omega/2k_B T_2) + \quad (20)$$
$$+ (k_x^2 + k_y^2)\Delta'(\omega)\alpha''(\omega + k_x V + \Omega) \coth(\hbar(\omega + k_x V + \Omega)/2k_B T_1) +$$
$$+ (k_x^2 + k_y^2)\Delta''(\omega)\alpha'(\omega + k_x V + \Omega) \coth(\hbar\omega/2k_B T_2)\}$$

$$\dot{Q} = \frac{\hbar}{4\pi^2} \int_{-\infty}^{+\infty} d\omega \int d^2k \frac{\omega}{k} \exp(-2k z_0) \cdot$$
$$\cdot \{k^2 \Delta''(\omega)\alpha''(\omega + k_x V)[\coth(\hbar\omega/2k_B T_2) - \coth(\hbar(\omega + k_x V)/2k_B T_1)] + \quad (21)$$
$$(k_x^2 + k_y^2)\Delta''(\omega)\alpha''(\omega + k_x V + \Omega)[\coth(\hbar\omega/2k_B T_2) - \coth(\hbar(\omega + k_x V + \Omega)/2k_B T_1)]\}$$

$$M_z = -\frac{\hbar}{4\pi^2} \int_{-\infty}^{+\infty} d\omega \int d^2k \frac{(k_x^2 + k_y^2)}{k} \exp(-2kz_0) \cdot \Delta''(\omega)\alpha''(\omega + k_x V + \Omega) \cdot \qquad (22)$$
$$\cdot \left[\coth(\hbar\omega/2k_B T_2) - \coth(\hbar(\omega + k_x V + \Omega)/2k_B T_1)\right]$$

One can see, that formulas (11),(15),(19) for $F_x$, formulas (12),(16),(20) for $F_z$, formulas (13),(17),(21) for $\dot{Q}$, and formulas (14),(18),(22) for $M_{x,y,z}$ are obtained one from another by a cyclic permutation $k_x^2 \to k_y^2, k_y^2 \to k^2, k^2 \to k_x^2$. In the case $\mathbf{\Omega} = 0, V \neq 0$ and $\mathbf{\Omega} \neq 0, V = 0$ Eqs. (11)—(22) are in complete correspondence to the results [4,8] and [11,12].

For a particle with magnetic polarizability $\alpha_m(\omega)$ and a surface with magnetic permeability $\mu(\omega)$, formulas (11)—(22) are valid with the replacements $\alpha(\omega) \to \alpha_m(\omega), \varepsilon(\omega) \to \mu(\omega)$, while in general case where the particle and surface have both electric and magnetic polarization, we have to take a sum of the corresponding results.

## Conclusions

Using a general background of the fluctuation electromagnetic theory, we have obtained closed nonrelativistic expressions for the friction/attraction forces, frictional torque and heating rate of a small spherical particle moving parallel to the surface at a constant velocity and simultaneously rotating in several directions relative to the velocity vector. Material properties of the particle and the surface are characterized by frequency-dependent electric(magnetic) polarizability and dielectric(magnetic) permeability. The temperatures of the particle and the surface are assumed to be arbitrary. In configurations without rotation or without translational motion the obtained formulas reduce to the known expressions in the corresponding cases.

## References

Figure 1a

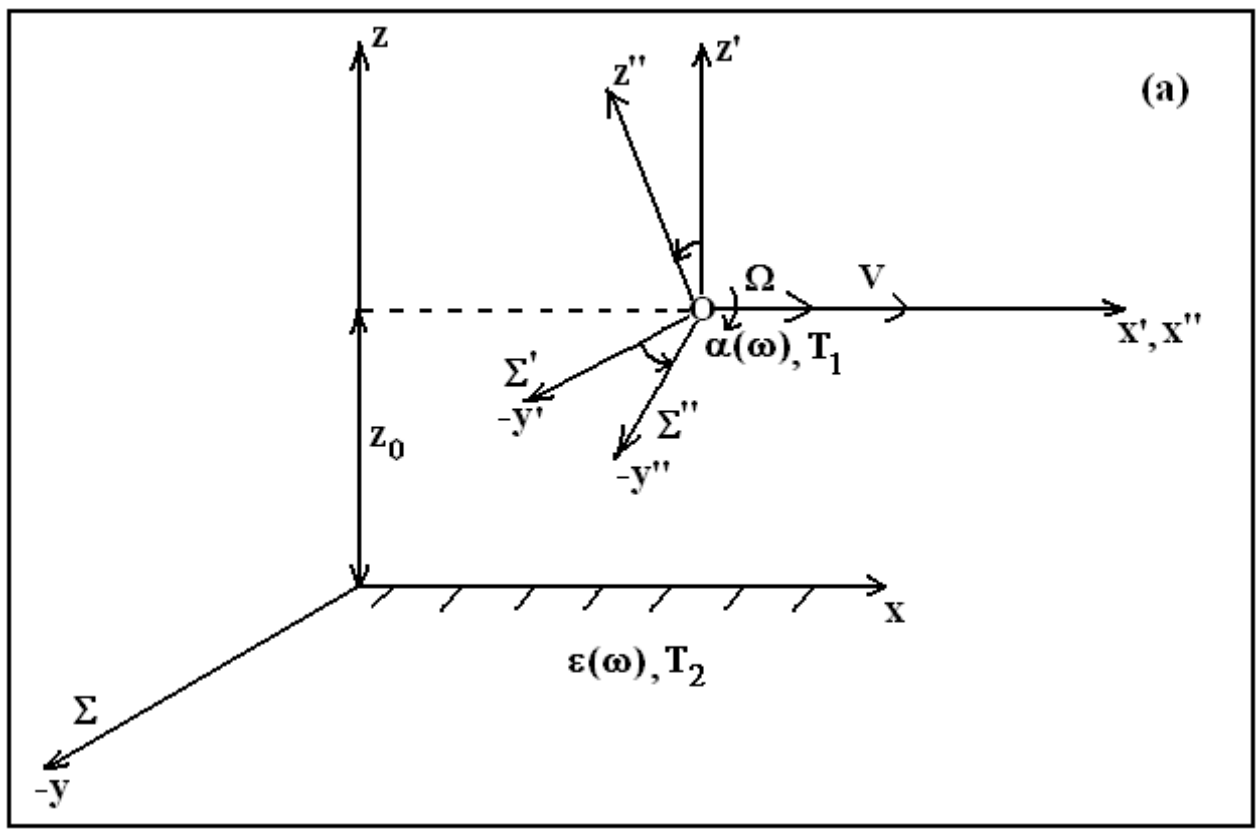

Figure 1b

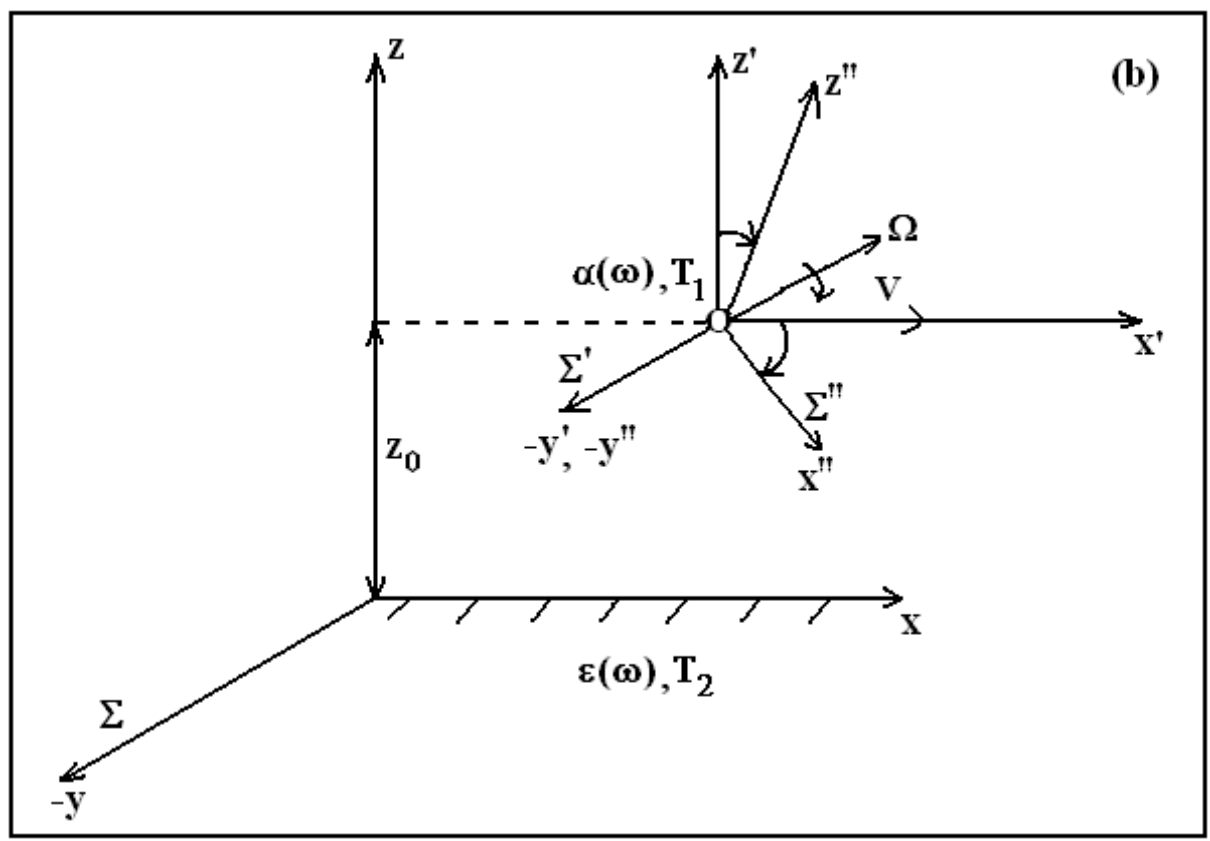

Figure 1c

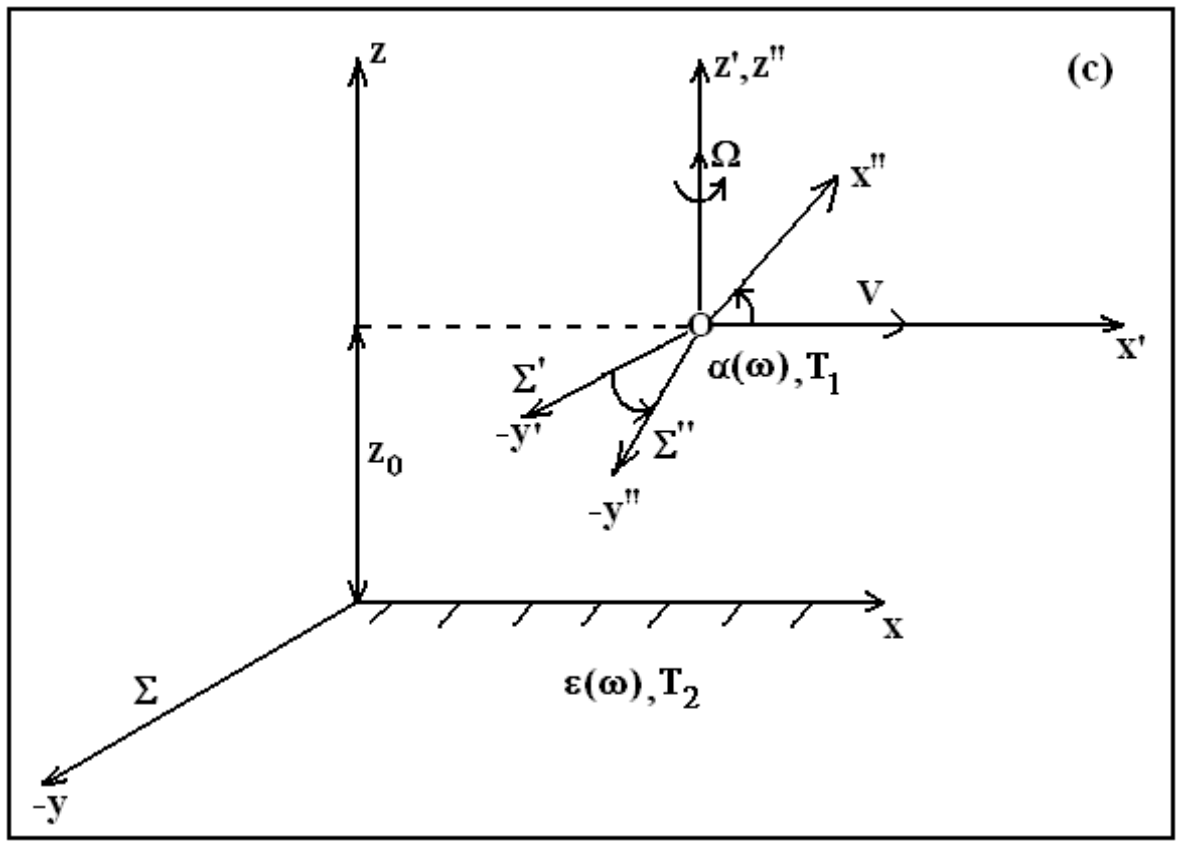

Fig.1(a,b,c) Geometrical configurations and coordinate systems.